\documentclass[useAMS,usenatbib]{mn2e}
\usepackage{graphicx}
\usepackage{dcolumn}
\usepackage{bm}

\title[Testing Dark Energy with ALPACA]{Testing Dark Energy with the Advanced Liquid-Mirror Probe of Asteroids, Cosmology and Astrophysics}
\author[Corasaniti, LoVerde, Crotts \& Blake]{Pier Stefano Corasaniti,$^{1,2}$ 
Marilena LoVerde,$^{1,3}$ Arlin Crotts$^{1,2}$ and Chris Blake$^{4}$\\
$^{1}$ISCAP, Columbia University, New York, NY 10027, USA\\
$^{2}$Department of Astronomy, Columbia University, New York, NY 10027, USA\\
$^{3}$Department of Physics, Columbia University, New York, NY 10027, USA \\
$^{4}$Department of Physics \& Astronomy, University of British Columbia, Vancouver, V6T 1Z1, Canada}
\begin{document}

\maketitle

\begin{abstract}
The Advanced Liquid-Mirror Probe of Asteroids, Cosmology and Astrophysics (ALPACA)
is a proposed 8-meter liquid mirror telescope 
surveying $\sim1000$ ${\rm deg}^2$ of the southern-hemisphere sky. It will be
a remarkably simple and inexpensive telescope, that nonetheless
will deliver a powerful sample of optical data for studying dark energy.
The bulk of the cosmological data consists of nightly, high signal-to-noise, 
multiband light curves of SN Ia. At the end of the three-years run ALPACA 
is expected to collect $\ga 100,000$ SNe Ia up to $z\sim1$. This will allow us to
reduce present systematic uncertainties affecting the standard-candle relation. 
The survey will also provide several other
datasets such as the detection of baryon acoustic oscillations in the matter power spectrum
and shear weak lensing measurements.
 In this preliminary analysis we forecast constraints on dark energy
parameters from SN Ia and baryon acoustic oscillations.
The combination of these two datasets will provide competitive constraints on the dark
energy parameters under minimal prior assumptions. Further studies
are needed to address the accuracy of weak lensing measurements.
\end{abstract}

\section{Introduction}

Over the past decade a picture of the Universe has emerged from
a number of cosmological observations which have
shed light on its geometry, matter content and clustering properties 
\citep{DeBe,Percival,Spergel,Tegmark}. 
Most astonishing amongst these findings is certainly the existence 
of ``dark energy'' which dominates the total
energy density of the Universe and which is responsible for its present state 
of accelerated expansion \citep{Riess,Perl}.

Despite the accurate measurements of the Cosmic Microwave Background
anisotropies, the SN Ia luminosity distance observations 
and the detailed mapping of the large scale matter distribution, 
the nature of this exotic component still remains unknown and
little progress has been made so far in this field, both at theoretical
and observational level.

The cosmological constant has been 
advocated as the simplest candidate,
since it is physically motivated representing the energy contribution of the vacuum. 
However we have no convincing explanation as to why the observed value is 
extremely small compared to particle physics expectations.
Alternative scenarios have flourished in the
recent past, although none
seems to provide a consistent particle physics formulation. 
For a recent review see Padmanabhan (2005). 

Supernova type Ia observations give the strongest evidence in favor of dark energy \citep{Riess04}, nonetheless
the limited number of measurements are still affected 
by large experimental and
systematic uncertainties which prevent further insights.
Consequently many questions about the properties of dark
energy still remain unanswered. 
For instance it would be crucial to know whether the dark
energy is time dependent or not, whether it clusters,
whether its effects are caused by an exotic form of matter or by
a different behavior of gravity on the large scales.
Recent data analysis have shown that current observations 
lack the necessary accuracy to address 
these very issues (see for instance Corasaniti et al. 2004 and reference 
therein).

Complementary to the SN Ia observations, several other tests of dark energy 
have been studied in a vast literature including baryon acoustic oscillations 
in the galaxy power spectrum \citep{Blake03,seo}, 
cluster number counts \citep{WellerB,Khoury}, weak lensing 
shear measurements \citep{Jain,Ludo}
just to mention a few. Improving the accuracy and reducing the systematics
is the necessary condition for these methods to be effective. 
However they are all limited to some extent by degeneracies between dark energy 
and the other cosmological parameters. It is therefore unlikely that one single 
experiment may reveal the nature of dark energy. Progress seems only possible 
through a synergy of these different tests.

The Advanced Liquid-mirror Probe of Asteroids, 
Cosmology and Astrophysics (ALPACA) is a proposal for the realization of
a ground based liquid-mirror telescope surveying a strip of sky of
the southern-hemisphere. The survey will collect a huge amount of
data divided in several datasets pertaining to SN Ia light curves, 
high redshift galaxies and weak lensing. 

In this paper we study the constraints
on dark energy from the estimated distribution of SN Ia observed during the
nominal duration of the survey, and from
the expected level of detection of the baryon acoustic oscillations in the galaxy power spectrum.
The paper is organized as follows. In Section~\ref{alpaca} we 
give an overview of the project. In Section~\ref{SNBAO}
we discuss datasets that the survey is expected to provide, in particular SN Ia observations,
baryon acoustic oscillations and weak lensing. 
In Section~\ref{fisher} we introduce the Fisher matrix
analysis and in Section~\ref{results} we present our results. Finally
in Section~\ref{conclu} we discuss our conclusions.

\section{ALPACA project}\label{alpaca}

ALPACA is an 8-meter liquid mirror 
telescope which will incorporate a $3^{\circ}$-wide 
imaging field and will be sited on Cerro Tololo (CTIO) in northern Chile. 
It is amazingly cost effective, delivering more information than virtually 
any competing survey instrument but at a small fraction of the cost. 
It will be a remarkably simple and inexpensive telescope that nonetheless 
will deliver powerful samples of optical data that are suitable for dark energy studies. 

The concept of large, liquid-mirror telescope (LMT) is fairly new, although LMTs have
successfully operated for over a decade (Cabanac, Borra \& Beauchemin 1998; 
Hickson \& Mulrooney 1998). Serious difficulties which compromised the
past optical performance of LMT have now been eliminated. The result of these advancements
in the liquid-mirror technology is seeing-dominated imaging. 
The next generation of LMT such as ALPACA can therefore compete 
with standard large optical telescopes for many projects. 

ALPACA imaging quality is expected to be limited only by the seeing
at CTIO, hence $\la 1^{''}$. The cosmological survey 
field will cover ${\rm 775}$ ${\rm deg^2}$ in five bands ($u,b,r,i,z$) 
with limiting nightly AB magnitudes in the range 23rd to 25th 
over a large fraction of the field. 
Four of the five photometric bands $u$ (310-410nm), $b$ (415-555nm), 
$r$ (560-745nm) and $i$ (750-1050nm) are spaced evenly in $log(\lambda)$ 
($\lambda$ being the wavelength), so that K-correction uncertainty vanishes at the special ratios 
of $(1+z)=1.25,1.58$ and $1.98$. A fifth band $z$ (950-1050nm) is especially useful 
for isolating high-redshift objects. An additional $100-200$ ${\rm deg^2}$ will be covered 
at lower sensitivity. ALPACA will observe the field passing directly overhead at CTIO, 
some $3^{\circ}$ across, of which $2.5^{\circ}$ are well suited for cosmological work. 
The focal plane's CCDs operate in drift-scan mode, each CCD 
reobserves the same strip of sky each night, 
making image subtraction and further analysis procedures stable, simple and redundant. 
We are also exploring installation of a multiobject spectrograph.
The nominal duration of the survey is 3 years, 
at which time the flux depth will approach the crowding limit. 
The total amount of data collected during the three years run is expected to be 
of several petabytes. 

Observations of SN Ia standard candle luminosity 
distance are expected to provide the bulk of the cosmological data, other datasets will 
include weak lensing, baryon oscillations, cluster number counts, strong-lens time delays, 
and cross-correlation multiple-waveband methods, such as ISW-correlation with CMB maps 
and X-ray galaxy clusters.

\section{Data products}\label{SNBAO}

\subsection{SN Ia standard candles}\label{sn}

Supernovae Type Ia light curves are related by one parameter family relation
between the time-width and the maximum-light peak \cite{Phillips93,Hamuy}. 
As consequence of this SN Ia are standardizable candles and can be used as distance indicators 
(Riess, Press \& Kirshner 1996; Perlmutter et al. 1997; Phillips 1999). 
Models of supernova have been intensely investigated for decades, 
but only in the past few years have three-dimensional numerical simulations 
provided a better understanding of the physical processes which take 
place during the explosive phase (see Hoflich et al. (2003) for a recent review).

The most accredited scenario consists of a carbon/oxygen (C/O) 
White Dwarf (WD) accreting material from an evolved companion star. 
As the WD reaches the Chandrasekar mass, the compressional heating triggers a 
thermo-nuclear explosion in the WD's core that propagates to the outer layers. 
\begin{table*}
\centering
\begin{minipage}{140mm}
\caption{The redshift distribution of SN Ia expected from
ALPACA, where $z$ corresponds to the center of the bin.}\label{sndist}
\begin{tabular}{|@{}l|ccccccccccc@{}|}
\hline
$z$ & 0.1 & 0.19 & 0.28 & 0.37 & 0.46 & 0.55 & 0.64 & 0.73 & 0.82 & 0.91& 1.0\\
\hline
${\rm N(z)}$ & 300 &900 & 4500 & 9000 & 13500 & 19800 & 12600 & 12600 & 6750 & 4038 & 2204\\
\hline
\end{tabular}
\end{minipage}
\end{table*}
However we still lack of a comprensive understanding of the fine details
of the explosive phase which has limited the possibility to 
make robust quantitative predictions for the SNe light curves and spectra. 
In fact there is evidence that a number of secondary effects 
are responsible for deviations of the SN light curves from
the one parameter family relation. These may depend on conditions prior to the explosion, 
such as the initial metallicity of the WD, the C/O ratio as function of 
the mass of the progenitors, and features of the explosive phase, such as
rotational support, magnetic fields, convection structure in the 
deflagration front, viewing angle, ejecta
asymmetries, effects of the companion on the ejecta, circumstellar interactions
just to mention a few. Some of these may be purely stochastic, while others may leave
clues in multiband lightcurves which will allow their induced variance to be
removed or reduced.
Several theoretical treatments have attempted to predict the effect
on light curves of these various intrinsic factors (Hoflich, Wheeler \& Thielemann 1998;
Dominguez \& Hoflich 2000; Mazzali et al. 2001; Pinto \& Eastman 2001; Timmes et al. 2003).
Unfortunately, they tend not to agree well on the size, sign or even the rough
nature of these effects. In addition extrinsic effects, such as varying extinction laws or
gravitational lensing may also influence the luminosity and other lightcurve
observable as well. 

The use of the brighter-slower relation has already allowed for a reduction of the dispersion 
in the magnitude-luminosity distance to $\sim 15\%$ r.m.s.
(Phillips 1993, 1999; Riess, Press $\&$ Kirshner 1996; Perlmutter et al. 1997;
Goldhaber et al. 2001). 
Several studies have shown that this scatter can be further reduced using color data. 
For instance using the CMAGIC method \citep{Wang03} to find when $B-V$ color reaches a certain 
level after maximum light and thereby inferring luminosity, 
the dispersion can be limited to $8-10\%$. This method still requires an accurate 
measurement of the time-width of the maximum-light peak. 
On the contrary using the $B-V$ color 12 days after maximum-light 
($\Delta C_{12}$) can reduce the scatter to $\sim 7\%$ or less \citep{Wang05} 
without the need of the maximum-light peak width. 

The ALPACA survey strategy is optimal for detecting a huge sample of supernovae, 
providing superlative quality dataset of nightly, high signal-to-noise, 
multiband SN Ia light curves. At the end of the three years run ALPACA 
is expected to observe $\ga 100,000$ supernovae distributed in ten equally 
spaced redshift bins in the range $0.2<z<1$, and several hundred 
low-z supernova ($z<0.2$). Hence a primary utility of the ALPACA dataset would be 
to improve the accuracy of the SN Ia standard candle relation and reduce
systematic effects. 
In Table~\ref{sndist} we list the expected SN Ia redshift distribution.\footnote{We thank Ben Johnsonn 
for computing the expected SN Ia redshift distribution.}

As discussed above, the theoretical uncertainties in the physics of SN Ia 
leave us without indication on how individual characteristics of the progenitors or
peculiar conditions during the explosive phase 
may systematically affect the standard-candle relation. 
If such effects exist we can only hope to identify them through the component analysis
of large sample of nightly, high S/N, multiband light curves.
Although it is not {\it a priori} obvious which observational features might
tie to luminosity, we might already have hints. For example, the CMAGIC technique (Wang et al. 2003)
uses intensively sampled multiband lightcurves to produce a color-based correction
to the maximum-light standard-candle relation. However there is still
more information in such light curves. As an example, B-band ``bump''
of additional luminosity above the observed, or linear $M_B$
versus $B-V$ evolution can add up to $\sim 0.5\,{\rm mag}$
to the B luminosity near maximum light, which indeed seems to be at least
roughly correlated with the SN peak luminosity. Such features
might become manifest as additional higher order components to the brighter-slower relation.
Using our high signal-to-noise, densely sampled, multicolor lightcurves,
we expect to be able to measure from the component analysis
even $4$th and $5$th-moments, and their possible correlations with luminosity.

Furthermore multiband lightcurve shape may be tied to the overall SN luminosity.
For instance, it has been suggested that under multiple scattering
conditions, light echoes can affect the effective luminosity and
color of SNe Ia events (Patat 2005). Single scattering has negligible
effect at the level of accuracy required in the SN Ia standard-candle relation, but
in the case of optical-thickness of order of unity, circumstellar echoes
may prove important even for SNe Ia near
maximum light, to the point of affecting stretch or color-based luminosity corrections significantly.
Such an echo will affect various bands differently, and may be or may not be associated
with high extinction, depending on the patchiness of the surrounding dust. 
These characteristics makes ALPACA nightly multiband lightcurves an optimal
benchmark to measure this effect.
 
In addition the large ALPACA dataset will allow us to search 
for possible subclasses of SN Ia and to the discovery of new
standard-candle relations at different redshifts.
Since we can already disentangle SN Ia 
from core-collapse SN via their multiband light curves alone (Johnson and Crotts 2005), 
we can simply rely on spectroscopy of a few representative members 
for characterization of each Ia subclass we discover 
(not necessarily for type classification of every SN). 
In fact hydrogen-rich supernova (Type II) can be distinguished
from the hydrogen-poor (Ia, Ib and Ic) through the UV deficit, since
UV radiation is blocked in Type I explosions
by metal lines.
Contamination from SN Ib/c can be particularly relevant 
at high redshift and affect the cosmological parameter estimation \cite{Hom}. 
Efficient photometric identification of Ib/c is also possible
using four-band photometry focused on the bluer bands and with modest
photometric accuracy \cite{Gal}. 
At the highest redshifts ($z\ga 1$) weak lensing from the intervening
mass distribution introduce a scatter of
about $10\%$ (with zero shift in mean), and even weak lensing maps can
correct only a small fraction of this effect (Dalal et al. 2003).
This scatter is much smaller at lower redshifts ($z<1$) and 
can eventually be inferred from the data (Amanullah, Mortsell \& Goobar 2003).
Due to the large number of SN Ia per redshift bin this effect
is negligible and averaged out.

A sample of tens or hundreds of SN Ia studied in this way might 
reveal the spectroscopic characteristics associated with these photometric 
properties. Furthermore, if we can also measure the redshifts of these 
SNe's host galaxies, we can derive accurate luminosities. 

Although ALPACA is conceived as a drift-scanned imaging telescope, it is not 
impossible to suggest how a spectrograph might be introduced at a later time. 
This instrument would also drift scan, this time through the reimaging system 
of the spectrograph onto the time-delay integration (hence drift-scanned) 
CCD detector at the back of the spectrograph. A spectrographic slit mask 
could also be scanned across the focus of the telescope in front of the 
spectrograph in a way which might deliver many spectra simultaneously. 
The implementation of this idea might also deliver spectra for many 
thousands of supernovae, and could 
study huge numbers of galaxies as well. 

Measuring the apparent 
brightness and host galaxy redshift for these supernovae would allow us 
to compute their luminosities to the level of 1\% or better relative to other 
SN within redshift intervals of $\Delta z \approx 0.1$ (even given the current uncertainty 
in the background cosmology).  
This will give us the opportunity 
to isolate samples of Type Ia SNe within redshift bins of $\Delta z\approx0.1$.
In one year's survey, each bin will contain thousand of SNe. 
Comparing these large samples of SNe of nearly identical redshifts
(therefore nearly equidistant), will allow us to seek for subclasses of SNe Ia within each redshift bin.
Such subclasses could be distinguished on the basis of 
temporal evolution in various bands (for example different 
expansion rates at the same peak luminosity would indicate different
temperature evolutions beyond those implied by $^{56}$Ni mass),
or from higher-order moments of the light curves. 
If the fractional contributions of the various subclasses vary with
redshift, we could then compensate for the effect. Furthermore,
the various subclasses will be checked against one other to identify
and correct for any systematic drifts within individual subclasses.

\subsection{Baryon Acoustic Oscillations}\label{bao} 
Baryon oscillations in the galaxy power spectrum have recently emerged as 
a promising probe of dark energy \cite{Blake03,seo}. 
The large-scale linear clustering pattern contains a 
series of small-amplitude, roughly sinusoidal, modulations in power of 
identical physical origin to the acoustic peaks observed in the CMB. 
These oscillations encode a characteristic scale, the sound horizon at 
the drag epoch, which can be used as a cosmological 
standard ruler in the low-redshift Universe. When applied in the 
tangential and radial directions, this method enables the measurement of 
the angular diameter distance $D_A(z)$ and Hubble parameter $H(z)$ in units of 
the sound horizon, over a series of redshift slices, which in turn may be 
used to place accurate constraints on dark energy models.

The preferred scale was recently identified in the clustering pattern of
Luminous Red Galaxies in the Sloan Digital Sky Survey \cite{Eis}, 
constituting an important validation of the technique. A power
spectrum analysis of the final 2-degree Field Galaxy Redshift Survey
produced consistent results (Cole et al. 2005).  
The baryon oscillations method is robust against systematic errors. 
The underlying structure of acoustic peaks can be modeled very accurately 
using the linear physics of the CMB. In the galaxy distribution there are 
modifying ``non-linear'' effects owing to redshift-space distortions, halo 
bias and non-linear growth of structure. However, these effects are 
confined to broad-band changes to the power spectrum and do not introduce 
preferred scales that could be confused with the early-universe sound 
horizon. Recent simulations have confirmed that the baryon oscillation 
signature is preserved with only minor degradation (Seo \& Eisenstein 2005, 
Angulo et al. 2005, Springel et al. 2005, White 2005). 
The acoustic signature may also be measured from photometric redshift 
surveys \cite{seo,Blake04}, although the 
smearing of radial information implies that only $D_A(z)$ can be determined 
with any confidence. However, given a sufficiently deep survey and 
accurate photometric redshifts, baryon oscillations may be detected with 
only $\sim1000$ ${\rm  deg}^2$ of imaging in broad redshift slices centered at $z=1$ and $z=3$. 
We use the technique of \cite{Blake04} to simulate the 
resulting measurements of $D_A(z=1)$ and $D_A(z=3)$. The expected galaxy number 
guarantees that the clustering measurements are limited by 
cosmic variance and not by shot noise ($nP = 3$ in the notation of Seo \& Eisenstein 2003). 
This is entirely consistent with the anticipated survey depth of ALPACA. 
We assume a survey area of $775$ ${\rm deg}^2$. For some galaxy classes, 
accurate photometric redshifts are possible with five broad bands, 
for Luminous Red Galaxies \cite{Padmanabhan} $\sigma_z=\sigma_0 (1+z)$ with $\sigma_0=0.03$ at $z<1$, 
due to the strong spectral break which implies rapidly changing colours with redshift. For other 
classes $\sigma_0 = 0.06$ \cite{Soto}, while good photometric performance should also 
be possible for $z=3$ galaxies using the Lyman break. For a recent analysis
see Ilber et al. (2006). We suppose that these 
accuracies are achievable over the ranges $0.5 < z < 1.5$ and $2.5 < z < 3.5$. 
(We omit $1.5 < z < 2.5$ given that photometric redshifts will not 
be accurate in this range in the absence of near-infra-red data). 
For comparison, we also display the equivalent measurements from 
a spectroscopic survey with the same configuration, using the 
method of Glazebrook \& Blake (2005). Our results are listed in Table~\ref{errphotoz}. 
An additional potential systematic error for an ALPACA large-scale structure 
survey is the survey geometry, which is a long, narrow stripe. 
Statistically, the number of measured Fourier modes is independent of the 
survey area. However, in order to robustly measure the clustering 
amplitude on a given scale, that scale must significantly exceed the 
minimum dimension of the survey.  Otherwise we risk an enhanced cosmic 
variance owing to our consequent inability to measure the underlying 
large-wavelength modes in that dimension. In our case, the stripe width 
($2.5$ ${\rm deg}$, equivalent to a projected $101$ ${\rm Mpc/h}$ 
at $z=1$ and $194$ ${\rm Mpc/h}$ at $z=3$) is comparable to some 
of the scales of interest for baryon oscillations ($20-100$ Mpc). 
Further simulations are required to quantify the level of risk for enhanced cosmic variance.

\subsection{Weak lensing and other datasets}
ALPACA will also provide a very deep imaging survey perfectly
suited for gravitational lensing studies. After three years
of the survey, the cumulative AB magnitude limit in the $r$-band
will be $\sim 27$, providing $\ga 40$ ${\rm gal/arcmin^2}$. The imaging
quality will be limited only by seeing at the CTIO site, thus
no worse than $1$ ${\rm arcsec}$. With this resolution ALPACA
will be capable of mapping the dark matter distribution from
cosmic shear measurements with unprecedent resolution.
Small scale resolution (on the arcmin scales) combined with measurements
up to $1^{\circ}$ are particularly important to constrain dark energy.
For a given amplitude of the scalar perturbations 
the balance between the non-linear scales and the large ones is sensitive
to the dark energy parameters \cite{Ludo}. A more 
powerful method is to use lensing ``tomography'' \cite{Hu99}
from different redshift bins. ALPACA's sensitivity ensures
that the noise coming from the intrinsic ellipticity of galaxies
is negligible at all scales, offering an optimal
measurement of gravitational lensing in the range of interest.
A known source of limitation arises from the correction of galaxy
shapes for Point Spread Function (PSF) effects. 
Foreground stars in the ALPACA stripes can be used
to determine the anisotropic PSF and correct the images 
with standard techniques (Kaiser, Squire \& Broadhurst 1995; Heymans et al. 2005).
In contrast the isotropic PSF tends to circularize the galaxy shape
and therefore to dilute the lensing signal. In such a case
a correction factor proportional to the seeing must be applied.
A seeing $\la 1''$ is necessary in order to keep the uncertainty of the shape
correction factor below $5\%$ \cite{Erben}. Improving the observational accuracy
is not the only complication, on the theoretical side we have a limited
knowledge of the growth of the non-linear structures. Using N-body simulations
it has been possible to predict the non-linear power spectrum up to $1$ ${\rm arcmin}$
with $5\%$ accuracy at most \cite{Smith}. 

An interesting characteristic of ALPACA is that the same region of the sky will
be sampled $\sim1000$ times. This is an entirely new regime of the PSF correction, which
has never been explored before. In fact the number of dithered images with
current lensing survey is $\sim 10$. The larger number of exposures could greatly help to better
determine the effect of systematics. We leave the analysis 
of ALPACA weak lensing observations to future studies.

The survey will measure the location and
collect photometric redshifts of $\sim 10^8$ galaxies over a large
range of redshifts, potentially leading to the identifications of many galaxy
clusters and groups. This will allow for an accurate number counting test, which 
in combination with existing measurements of Cosmic Microwave Background 
power spectrum can provide complementary constraints on the dark energy
parameters \citep{Khoury}. Another use of ALPACA would be to measure
the cross-correlation between the survey and
the CMB maps for measuring the Integrated Sachs-Wolfe (ISW) effect (Crittenden
 \& Turok 1996).
Although due to the limited sky coverage of the survey these measurements may not be able 
to provide competitive constraints on dark energy parameters \cite{Levon}.
The survey will also provide a large sample of quasars for Alcock-Paczynski
test (Calvao, De Mello Neto \& Waga 2002), as well as strong-lens time delays (for
10-20 QSOs) \cite{Shapiro}. 
 
\section{Fisher Matrix Analysis}\label{fisher}
We use the Fisher matrix approach to forecast constraints on dark energy parameters from
ALPACA observations of SN Ia and acoustic baryon oscillations.
The Fisher matrix provides a simple and practical method for estimating
parameter errors. It gives a local approximation to the likelihood 
surface about a fiducial model for a set of independent observations (Tegmark, Taylor $\&$ Heavens 1997). 
Formally the Fisher matrix reads as
\begin{equation}
F_{\mu\nu}=\sum_i \frac{1}{\sigma^2{(\mathcal{O}_i)}}\frac{d\mathcal{O}_i}{d\theta_\mu}
\frac{d\mathcal{O}_i}{d\theta_\nu},\label{fish}
\end{equation}
where $\theta_\mu$ are the model parameters, $\mathcal{O}_i$
are the observations of the quantity $\mathcal{O}$ and $\sigma{(\mathcal{O}_i)}$ the experimental errors. 
The derivatives in Eq.~(\ref{fish})
are computed at the fiducial model parameter values. 
The covariance matrix is given by the inverse of Fisher matrix, $C=F^{-1}$,
with the one sigma errors on the model parameters given by the diagonal components,
$C_{\mu\mu}=\sigma^2_{\theta_\mu}$.
In case of multiple independent datasets the combined constraints
can be inferred by simply adding the Fisher matrices of each experiment.

Supernova Ia observations measure the luminosity distance $d_L$ through 
the standard-candle relation,
\begin{equation}
m(z)=\mathcal{M}+5\log{\mathcal{D}_L(z)},
\end{equation}
where $\mathcal{M}\equiv M-5\log{H_0}+25$ is the ``Hubble-constant-free'' absolute
magnitude and $\mathcal{D}_L(z)\equiv H_0 d_L(z)$ is the ``Hubble-constant-free''
luminosity distance. In flat universe the luminosity distance is given by
\begin{equation}
d_L(z)=(1+z)\int_0^z \frac{dz'}{H(z')},
\end{equation}
with
\begin{equation}
H(z)=H_0 \left[(1-\Omega_{m})f_{DE}(z)+\Omega_m(1+z)^3\right]^{1/2},
\end{equation}
where $\Omega_m$ is the dark matter
energy density and 
\begin{equation}
f_{DE}(z)=exp\left\{3\int_0^z\frac{1+w(z')}{1+z'}dz'\right\}\label{fde}
\end{equation}
with $w(z)$ being the dark energy equation of state. 
Several parameterizations have been extensively studied in the literature, for simplicity
we limit our analysis to dark energy models
parametrized by \citep{Polarski,Linder}:
\begin{equation}
w(z)=w_0+w_a\frac{z}{1+z}.
\end{equation}

We have a three-parameter space consisting of
$\Omega_m$, $w_0$ and $w_a$. In addition the magnitude-luminosity distance relation depends on the off-set
parameter $\mathcal{M}$ which we marginalize over. For the fiducial cosmology we adopt
a model with $\Omega_m=0.3$, $w_0=-1$ and $w_a=0$. 
We compute the SN Ia Fisher matrix 
using Eq.~(\ref{fish}) and the derivatives of $d_L(z)$ with respect to the model parameters
using double-finite differences about the fiducial model. 

We account for the uncertainty of the observed magnitude in each redshift bin,
using the standard approach of adding in quadrature the statistical and the irreducible 
systematic errors (see for instance Kim et al. (2004)),
\begin{equation}
\sigma^2(z_i)=\frac{\sigma_{stat}^2}{N_i}+\sigma_{sys}^2,\label{snerr}
\end{equation}
with $\sigma_{stat}=0.15$ and $N_i$ being the number of SN Ia in the $i$-th bin.
Assessing the amplitude of the expected level of systematic uncertainty $\sigma_{sys}$ 
is more difficult. Because observations at higher redshift are more challenging 
the systematic error should increase, this increase is usually assumed to vary linearly with the redshift,
\begin{equation}
\sigma_{sys}=\delta{m}\frac{z}{z_{max}},
\end{equation}
where $\delta{m}$ is the amplitude of the systematics at the maximum 
redshift bin of the survey $z_{max}$. As we have discussed in Section~\ref{sn},
we still lack robust quantitative prediction for SN Ia light curves and spectra.
Although results from numerical simulations remain controversial, they seem
to suggest that nightly high signal-to-noise
multiband SN Ia light curves may suffice to reduce systematic uncertainties to few
$\%$ level (see for instance Hoflich, Wheeler \& Thielemann 1998). 
The use of multiple methods on a vast SN Ia sample 
such as the one provided by ALPACA is likely to reduce the level of systematics 
to $5-7\%$ or better. Constraints from future
SN Ia survey usually assume a more optimistic $2\%$ uncertainty.
We forecast constraints for both $\delta{m}=0.05$ and $0.02$. 

As discussed in Section \ref{bao}, baryon acoustic oscillations 
in the matter power spectrum provide
a standard ruler which can be used to infer cosmological distance measurements.
Oscillations in the tangential component of the matter power spectrum
give the comoving angular diameter distance
at the redshift slice of the survey in units of the sound horizon,
\begin{equation}
x(z)=(1+z)\frac{d_A(z)}{s},
\end{equation}
where $d_A(z)=d_L(z)/(1+z)^2$ is the angular diameter distance and
the sound horizon $s$ is given by
\begin{equation}
s=\int_0^{a_{drag}} \frac{c_s}{\sqrt{\Omega_m a+\Omega_r}}da,
\end{equation}
where $a_{drag}$ is the value of scale factor at the epoch of baryon-drag (baryons decoupling) and
$\Omega_r$ is the radiation energy density.
The sound speed of the barotropic fluid is
\begin{equation}
c_s^2=\frac{1}{3}\frac{1}{1+\frac{3}{4}\frac{\Omega_b}{\Omega_r}a}.
\end{equation}
Oscillations in the radial component give
the Hubble rate in units of the sound horizon,
\begin{equation}
x'(z)=\frac{H^{-1}}{s}.
\end{equation}
Because of the dependence on the sound horizon,
the parameter space is extended to $\Omega_b$ and the Hubble constant $h=H_0/100$.
Again we use Eq.~(\ref{fish}) to compute the Fisher matrix components. The expected errors
on $x$ and $x'$ from ALPACA measurements are listed in Table~\ref{errphotoz}.

\begin{table}
\caption{Expected uncertainties on $x=(1+z)d_A(z)/s$ and $x'=H^{-1}(z)/s$ in different redshift bins,
for different redshift measurements.}\label{errphotoz}
\begin{tabular}{|c|c|c|c|c|}
\hline
$z$ & $\sigma_{x}$ ($\sigma_0=0.03$) & $\sigma_x$ (spect-z) & $\sigma_{x'}$ (spect-z)\\
\hline
1.0 &  0.039 & 0.013     &   0.022\\
3.0 &  0.020 & 0.009     &   0.016\\
\hline
\end{tabular}
\end{table}

\section{Results}\label{results}
We now discuss the results of the Fisher matrix analysis. We list in Table~\ref{dm0.05} and
Table~\ref{dm0.02} the $1\sigma$ limits on $\Omega_m$, $w_0$ amd $w_a$ from SN Ia with $\delta{m}=0.05$
and $0.02$ systematic uncertainty respectively. We also quote the errors inferred from assuming the standard
prior $\sigma_{\Omega_m}=0.03$ and compare with those obtained combining baryon acoustic oscillations.

Supernova data alone poorly constrain the dark energy parameters due to the degeneracy in $\Omega_m$,
for instance for $\delta{m}=0.05$, we find $\sigma_{\Omega_m}=0.62$, $\sigma_{w_0}=0.49$ 
and $w_a$ unbounded. These errors reduce by nearly a factor 
$2$ for the case $\delta{m}=0.02$ (Table~\ref{dm0.02}). 
On the contrary assuming the standard $\Omega_m$ prior we obtain $\sigma_{w_0}=0.2$
and $\sigma_{w_a}=1.1$ for $\delta{m}=0.05$, and $\sigma_{w_0}=0.11$ and $\sigma_{w_a}=0.53$ for $\delta{m}=0.02$.

\begin{figure}
\begin{center}
\includegraphics[width=80mm]{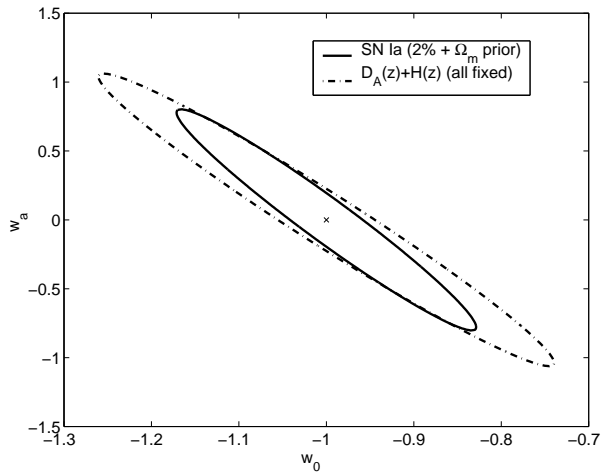}
\caption{1-$\sigma$ Fisher matrix contours on $w_0-w_a$. The solid line
corresponds to SN Ia with $2\%$ systematic uncertainty 
and $3\%$ Gaussian prior on $\Omega_m$, while the dash-dot line corresponds to
BAO measurements with spectroscopic redshift accuracy assuming
perfect knowledge of $\Omega_m$ and $h$.}\label{fig1}
\end{center}
\end{figure}

Constraints on dark energy from baryon acoustic oscillations are also limited by degeneracies with other cosmological
parameters, in particular $\Omega_m$ and $h$. In addition there is a further uncertainty due to the dependence on 
$\Omega_b$ which determines the size of the sound horizon at decoupling. Assuming these parameters to be perfectly
known we find $\sigma_{w_0}=0.17$ and $\sigma_{w_a}=0.70$ for spectroscopic redshift measurements. 
In figure~\ref{fig1} we plot the $1\sigma$ contours in the $w_0-w_a$ plane from SN Ia (solid line) and baryon
oscillations (dash-dot line).

We are particularly interested in the SN and BAO combined constraints, so as to determine how the combination 
of the baryon acoustic oscillations can realistically improve the dark energy limits from SN Ia under
minimal prior assumptions. In such a case we only assume a Gaussian prior on the baryon density, 
$\sigma_{\Omega_b}=0.006$ consistently with WMAP/BBN results (Spergel et al. 2003; Hansen et al. 2002) and $h$ to be fixed.

In the case of photometric redshifts, the combination of SN Ia and BAO
gives $\sigma_{\Omega_m}=0.01$, $\sigma_{w_0}=0.19$ and $\sigma_{w_a}=0.96$ for $\delta{m}=0.05$, and
$\sigma_{\Omega_m}=0.01$, $\sigma_{w_0}=0.11$ and $\sigma_{w_a}=0.52$ for $\delta{m}=0.02$.
The use of spectroscopic redshifts leads to slightly better constraints, for instance we find
$\sigma_{\Omega_m}=0.003$, $\sigma_{w_0}=0.14$ and $\sigma_{w_a}=0.58$ for $\delta{m}=0.05$,
and $\sigma_{\Omega_m}=0.002$, $\sigma_{w_0}=0.09$ and $\sigma_{w_a}=0.43$ for $\delta{m}=0.02$.
The level of detection of baryon acoustic oscillations as expected for ALPACA 
will provide poor information on dark energy beyond that which comes from better constraining
$\Omega_m$ to the level expected from future CMB experiments. In Table~\ref{planck} we list
the expected parameter errors from combining SN and BAO with CMB limits from Planck, 
the inferred constraints are compatible with those obtained in (Linder 2005).
Including the shear weak lensing can certainly improve these limits, 
we leave this to a future study. 
 
\begin{table*}
\begin{minipage}{140mm}
\caption{1-$\sigma$ uncertainties on $\Omega_m,w_0$ and $w_a$ 
from SN Ia assuming $5\%$ systematic error with Gaussian $\Omega_m$ prior 
and with different combinations of BAO detections.}\label{dm0.05}
\begin{tabular}{|c|c|c|c|c|c|}
\hline
 & SN Ia ($\delta{m}=0.05$) & $+\sigma_{\Omega_m}=0.03$ & $+D_A(z)$ ($\sigma_0=0.03$)& $+D_A(z)+H(z)$ (spect-z)\\
\hline
$\sigma_{\Omega_m}$ & 0.62 &   -  & 0.01     & 0.003 \\
$\sigma_{w_0}$ &      0.49 &  0.2 & 0.19     & 0.14  \\
$\sigma_{w_a}$ &       -   &  1.1 & 0.96     & 0.58  \\
\hline
\end{tabular}
\end{minipage}
\end{table*}

\begin{table*}
\begin{minipage}{140mm}
\caption{As in Table~\ref{dm0.05} with $2\%$ SN Ia systematic error.}\label{dm0.02}
\begin{tabular}{|c|c|c|c|c|c|}
\hline
 & SN Ia ($\delta{m}=0.02$) & $+\sigma_{\Omega_m}=0.03$ & $+D_A(z)$ ($\sigma_0=0.03$)& $+D_A(z)+H(z)$ (spect-z)\\
\hline
$\sigma_{\Omega_m}$ & 0.30 &   -   & 0.01    &  0.002 \\
$\sigma_{w_0}$ &      0.20 &  0.11 & 0.11     &  0.09 \\
$\sigma_{w_a}$ &       -   &  0.53 & 0.52     &  0.43  \\
\hline
\end{tabular}
\end{minipage}
\end{table*}

\begin{table*}
\begin{minipage}{140mm}
\caption{As in Table~\ref{dm0.02} with Planck.}\label{planck}
\begin{tabular}{|c|c|}
\hline
 & SN Ia + BAO (spect-z) + Planck\\
\hline
$\sigma_{\Omega_m}$ &  0.001 \\
$\sigma_{w_0}$ &  0.07 \\
$\sigma_{w_a}$ &  0.31  \\
\hline
\end{tabular}
\end{minipage}
\end{table*}

\section{Conclusions}\label{conclu}
Uncovering the nature of dark energy is the new challenge
of modern observational cosmology. Over the upcoming years 
lots of effort will be dedicated to improving luminosity 
distance measurements of SN Ia standard candles and extend observations
to complementary dark energy tests, such as baryon acoustic oscillations and
cosmic shear weak lensing. In order to accomplish such a program 
several experiments have been proposed both from ground and space.

Here we have presented the ALPACA survey and described 
the main datasets that the project is expected to provide.
The bulk of the data is represented by high signal-to-noise, multiband,
nightly SN Ia light curves. At the end of the three year run
ALPACA is expected to observe $\ga 100,000$ supernovae distributed in
the range $0.2<z<1$ and several hundreds at lower redshifts. This huge
dataset will be useful for finding possible subclass of SN Ia at different redshifts and
correlations in the multiband light curves beyond the brighter-slower relation. 
This will allow for calibration of the standard-candle relation and reduction of the systematic 
uncertainties.

In this paper we have forecasted limits on dark energy parameters 
from the expected redshift distribution of SN Ia and the 
detection of baryon acoustic oscillations in the matter power spectrum. 
Although the accuracy of BAO measurements
cannot compete with dedicated surveys such as KAOS \cite{KAOS} or 
LSST \cite{LSST}, combining them with SN Ia data will
provide competitive constraints under minimal prior assumptions.
ALPACA will also deliver several other datasets and will be particularly suitable for
weak lensing measurements. 

ALPACA stakes out a regime in survey parameter space that is unique
and useful, the intensive nightly, multiband monitoring of fields
of order 1000 ${\rm deg}^2$. While in principle other surveys such
as LSST, Pan-STARRS, JDEM or DES are instrumented for such 
intensive monitoring, in practice they are driven by other motivations
such as all-sky transient monitoring or weak-lensing mapping, 
devoting much less time than ALPACA to the 1000 ${\rm deg}^2$ goal.
Indeed, using these like ALPACA would be inefficient, since ALPACA
is extremely economical for this application compared to one of these
more flexible instruments.
Liquid-mirror techniques have now reached such a high level of quality
that liquid-mirror telescopes may be used for astrophysics and cosmology.
The ALPACA project will take advantage of this and 
provide high quality data at a fraction of the cost
of standard 8-meter telescopes or space missions. The vast possibilities of 
the ALPACA survey are still not fully explored 
and further investigation is needed.

\section*{Acknowledgments}
We are grateful to all members of the ALPACA collaboration
and we are particularly thankful to David Branch, Paul Hickson, Ben Johnson, Ken Lanzetta, Nick Suntzeff,
Jun Zhang, Yun Wang and Ludovic van Waerbeke for useful discussions and suggestions. 
P.S.C. is supported by Columbia Academic Quality Fund.

\end{document}